\documentclass[%
 reprint,
 superscriptaddress,
 amsmath,amssymb,
 aps,
]{revtex4-2}

\usepackage{graphicx}
\usepackage{dcolumn}
\usepackage{bm}
\usepackage{multirow}
\usepackage[T1]{fontenc}
\usepackage{enumitem}
\usepackage{mathtools}
\usepackage{xcolor}

\begin{document}


\title{Agent-based neutral competition in two-community networks
}

\author{Kota Ishida}
\email{kota_ishida@u.yale-nus.edu.sg}
\affiliation{Division of Science, Yale-NUS College, 16 College Avenue West, \#01-220 Singapore 138527, Singapore}

\author{Beata Oborny}%
\email{beata.oborny@ttk.elte.hu}
\affiliation{Institute of Biology, Lor\'and E\"otv\"os University, P\'azm\'any P\'eter s\'et\'any 1/C, H-1117 Budapest, Hungary}
\altaffiliation[Also at ]{Institute of Evolution, Centre for Ecological Research, Hungarian Academy of Sciences, Budapest, Hungary}

\author{Michael T.~Gastner}
\email{michael.gastner@yale-nus.edu.sg}
\affiliation{Division of Science, Yale-NUS College, 16 College Avenue West, \#01-220 Singapore 138527, Singapore}

\date{\today}

\begin{abstract}
Competition between alternative states is an essential process in social and biological networks.
Neutral competition can be represented by an unbiased random drift process in which the states of vertices (e.g., opinions, genotypes, or species) in a network are updated by repeatedly selecting two connected vertices.
One of these vertices copies the state of the selected neighbor.
Such updates are repeated until all vertices are in the same ``consensus'' state.
There is no unique rule for selecting the vertex pair to be updated.
Real-world processes comprise three limiting factors that can influence the selected edge and the direction of spread: (1) the rate at which a vertex sends a state to its neighbors, (2) the rate at which a state is received by a neighbor, and (3) the rate at which a state can be exchanged through a connecting edge.
We investigate how these three limitations influence neutral competition in networks with two communities generated by a stochastic block model.
By using Monte Carlo
simulations,
we show how the community structure and update rule determine the states' success probabilities and the time until a consensus is reached.
We present a heterogeneous mean-field theory that agrees well with the Monte Carlo simulations.
The effectiveness of the heterogeneous mean-field theory implies that quantitative predictions about the consensus are possible even if empirical data (e.g., from ecological fieldwork or observations of social interactions) do not allow a complete reconstruction of all edges in the network.
\end{abstract}

\maketitle


\section{\label{sec:introduction}
Introduction}

Numerous social and biological phenomena in complex networks can be modeled as dynamic processes in which vertices update their states by copying their neighbors.
In social networks, individuals (represented by vertices) tend to adopt the opinions, beliefs, or cultural traits of their peers with whom they are connected~\cite{bentley_random_2004, mesoudi_random_2009}.
In biological settings, the vertices can represent individuals or places inhabited by individuals.
For example, in an ecological habitat network, patches (vertices) are colonized by species from connected patches~\cite{hanski_metapopulation_1999, economo_biodiversity_2011, borthagaray_chapter_2015, moreljournel_its_2019, thompson_process-based_2020}.
Unbiased random drift processes are benchmark models in which the probability of a copying event is independent of the state of the vertex~\cite{hubbell_unified_2001, hahn_drift_2003}.
In a genetic model, this feature can be interpreted as ascribing equal fitness to all states.
In an ecological context, a random drift process assumes that none of the species is a stronger competitor within a habitat patch (a vertex) than any other.
A state can only gain more influence by occupying more influential positions in the network.
Although a random drift process is a highly simplified representation of real-world dynamics, it can serve as a null model for investigating positional influence.

The voter model is a paradigmatic example of a random drift process~\cite{clifford_model_1973, holley_ergodic_1975}.
It aims to model the spread of opinions in a social network.
The voter model also has applications in other contexts (e.g., spread of languages~\cite{castello_ordering_2006}, competition between species~\cite{kordzakhia_two-species_2005}, and genetic drifts~\cite{antal_evolutionary_2006}).
In the simplest versions of the voter model, each vertex is in one of two possible states: ``red'' or ``blue.''
Vertices repeatedly update their states by copying the state of a random neighbor.
Three conditions must be met for a successful copying event.
(1)~A vertex must send information about its state.
(2)~A connected vertex must be ready to update its state.
(3)~A connection between the sender and recipient must be active.
In an ecological setting, the equivalent conditions are as follows:
(1)~The species must produce colonizers (e.g., seeds) in the sender patch.
(2)~These colonizers must be able to establish their lives in the recipient patch.
(3)~The colonizers must find a way from the sender into the recipient patch (e.g., through a suitable ecological corridor). 
We investigated random drift processes in which one of the three conditions is the limiting factor in the spread of the states.
\begin{itemize}
  \item In a sender-limited process (SLP), all edges are permanently open, and the recipient immediately updates its state. However, vertices attempt to spread their states across randomly selected edges at a finite rate only.
  \item In a recipient-limited process (RLP), all vertices constantly attempt to spread their states, and all edges are permanently open.
  However, vertices only copy the states of their neighbors at a finite rate.
  \item In a connection-limited process (CLP), vertices attempt to colonize their neighbors at an infinite rate, and they update their state immediately after receiving a copy of a neighbor.
  However, the edges are only open for transmission at a finite rate.
\end{itemize}

We assume that the system is homogeneous in the sense that all vertices are equally active senders, they are all equally active recipients, and all edges are open for transmission with equal rates.
Under these assumptions, the conventional voter model corresponds to the RLP.
Previous studies have termed the SLP as a ``reverse voter model''~\cite{castellano_effect_2005} or an ``invasion process''~\cite{sood_voter_2008}.
Others have referred to the SLP as a ``birth--death process,'' the RLP as a ``death--birth process,'' and the CLP as ``link dynamics''~\cite{tan_when_2014}.

Early research on random drift models assumed that the network was either a complete graph~\cite{moran_random_1958} or a regular lattice~\cite{clifford_model_1973}.
In these networks, as well as in any other regular network (i.e., a network in which each vertex has the same number of neighbors), no difference exists between the SLP, RLP, and CLP; every vertex is equally likely to be a sender and equally likely to be a recipient.
In irregular networks, this symmetry is broken.
The resulting differences between the update rules have been discussed in the context of the voter model~\cite{castellano_effect_2005, sood_voter_2008}, evolutionary dynamics~\cite{antal_evolutionary_2006}, and game-theoretic settings~\cite{lieberman_evolutionary_2005, ohtsuki_evolutionary_2006, ohtsuki_simple_2006}.

The purpose of this study is to demonstrate the differences between the SLP, RLP, and CLP in networks with a community structure.
A community is conventionally defined as a subnetwork that contains a significantly higher number of edges than predicted by a null model (e.g., an Erd\H{o}s--R\'enyi graph with the same mean degree as the investigated network).
In this study, we define communities more broadly as subnetworks with different (i.e., higher or lower) densities than the corresponding Erd\H{o}s--R\'enyi graph; thus, we also cover core-periphery structures and disassortative topologies under the umbrella of community-structured networks.
We generated a community structure using a stochastic block model, which is a standard method for generating communities in networks~\cite{holland_stochastic_1983, abbe_community_2018}.
By considering networks with two communities, we demonstrate that, depending on the network structure and the factor that limits the transmission of a state, an initial minority state can become the more probable winner in the competition.
We also compare the amount of time required to reach a consensus (i.e., a condition in which all vertices are in the same state) in the three processes (SLP, RLP, and CLP).

In Sec.~\ref{sec:background}, we briefly review the relevant literature on random drift processes in modular networks.
In Sec.~\ref{sec:models}, we define our notation for the stochastic block model and specify how we implemented the random drift processes as agent-based models.
We present the heterogeneous mean-field theory of two-community random drift processes in Sec.~\ref{sec:hetero_mf}.
In Sec.\ \ref{sec:prob_red_cons}, we derive the probability with which one of the states becomes a consensus.
In Sec.~\ref{sec:mean_cons_time}, we compare the mean consensus times in the SLP, RLP, and CLP.
We conclude by discussing the implications of our findings in Sec.~\ref{sec:discussion}.

\section{\label{sec:background}Background}

Random drift models have a long history in mathematical biology, usually under the name ``neutral models.''
This name indicates that competing partners are assumed to be equally strong.
Since the introduction of the neutral theory of evolution~\cite{kimura_evolutionary_1968}, such neutral models have been intensively applied in the field of population genetics.
In this context, the ``colors'' (red vs.~blue or more colors) represent genotypes within a population (i.e., within a single species).
Another typical field of application is ecology~\cite{hubbell_unified_2001}, in which the ``colors'' mark different species living together in a community.

In both fields, neutral models are important references for the study of competition between genotypes/species (in genetics/ecology).
The agents can represent individuals or places occupied by individuals belonging to different species~\cite{czaran_spatiotemporal_1998}.
The earliest models typically used mean-field approximations~\cite{moran_random_1958}, assuming perfect mixing in the whole population; nevertheless, early literature also highlighted the idea of considering the structure of interactions between agents.
For example, the classic model of Wright~\cite{wright_isolation_1943} represents genetic drift in a population that is spatially divided into subpopulations (demes), which inhabit discrete habitat patches (``islands'').
Members of the demes mostly breed within themselves, except for a few migrants drawn from the rest of the population.
Numerous models later explicitly considered the interaction structure, representing it by a network.
These studies investigated the effect of the network topology on the dynamics of competition between genotypes/species (e.g.,~\cite{lieberman_evolutionary_2005, antal_evolutionary_2006, constable_population_2014}). 
Many of these models relaxed the assumption of neutrality (i.e.,\ allowing unequal competitive strengths), further extended the voter model by including more than two ``colors'' (beyond red and blue), and permitted new colors to enter the system (by mutation in the genetic models and by speciation or immigration from an external species pool in the ecological settings).

A large field of research, which aims at integrating local (patch-level) and regional species dynamics, is metacommunity ecology~\cite{hanski_metapopulation_1999}. 
Metacommunity models usually relax some of the aforementioned restrictive assumptions and assume interaction rules between species that are more complex than those in the voter model.
Nevertheless, we believe that the voter model, because of its simplicity, is an important baseline model for the dynamics of competition in metacommunities.

In the context of opinion formation, vertices typically represent individuals, and the network structure reflects social relations (e.g., acquaintances).
Some previous studies have investigated the dynamics of the voter model on a special type of modular network structure: two-clique networks (i.e., networks in which both communities are complete subgraphs).
Sood et al.~\cite{sood_voter_2008} found that, if the number of edges between the cliques is small, the opinion dynamics in the two cliques can be approximated by two independent diffusion processes.
Conversely, when the inter-clique connectivity is high, the average opinions in the two cliques quickly become equal and remain coupled until a consensus is reached.
Later studies showed how this finding can be explained by the theory of coalescing random walks~\cite{masuda_voter_2014} and a heterogeneous mean-field theory~\cite{gastner_voter_2019}.
Recent work has also investigated how the dynamics of the two-clique voter model change if the cliques are influenced by opposing external news sources~\cite{bhat_nonuniversal_2019, bhat_polarization_2020}.

Only a few previous studies have analyzed the voter model on general two-community networks in which the communities are not cliques.
These studies focused on the question of suitable representations of the underlying Markov chain~\cite{lamarche-perrin_information_2016}.
Numerical results for the consensus time~\cite{banisch_markov_2015} and the success probability of a single mutant~\cite{masuda_evolutionary_2009} have also been presented.
Our study goes beyond previous research by examining the process from various initial conditions and under three kinds of limitations: when the interaction rates between adjacent vertices are limited by the ability of vertices to send (SLP model) or receive information (RLP) or by the capacity to transmit information across an edge (CLP).

\section{\label{sec:models}Model}

We generated networks using the stochastic block model, which is a canonical model for modular networks that can generate various community structures~\cite{holland_stochastic_1983}.
In the special case of a two-community network, the stochastic block model takes the following data as input.
First, we partition $N$ vertices into two disjoint communities, $C_1$ and $C_2$.
The number of vertices in $C_i$ is denoted as $N_i$.
Between any two vertices in the same community $C_i$, a link exists with probability $P_i$.
The number of edges between $C_1$ and $C_2$, randomly selected from all $N_1 N_2$ pairs of vertices, is denoted by $X$.
In our parameterization of the stochastic block model, $X$ is a fixed deterministic value; thus, $X \geq 1$ guarantees that there are connections between the communities.
We discarded any networks in which communities were internally disconnected to ensure that a consensus could be reached.

\begin{figure}
\includegraphics[width=\columnwidth]{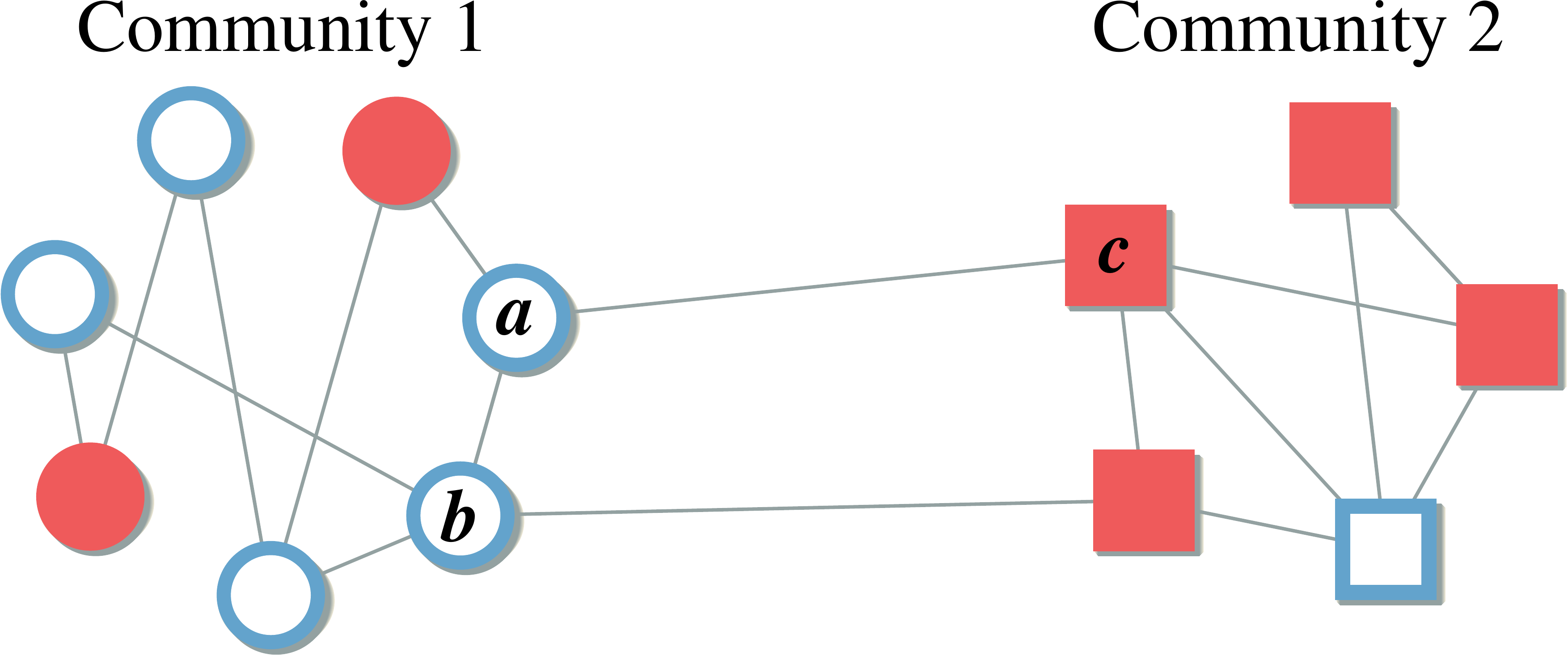}
\caption{\label{fig:netw_with_2_communities}
Illustrative network with two communities (circles vs.\ squares).
Each vertex has one of two possible states: red (closed symbols) or blue (open symbols).
In the SLP and RLP, we pick a random vertex (e.g., the blue vertex $a$) and a neighbor of that vertex.
If the chosen neighbor is the blue vertex $b$, $a$ maintains its current state.
If, instead, the chosen neighbor is the red vertex $c$, then either $c$ becomes blue (SLP), or $a$ becomes red (RLP).
In the CLP, we pick a random edge instead of a random vertex.
If the selected edge is $(a, c)$, either $a$ becomes red (with probability $1/2$) or $c$ becomes blue (also with probability $1/2$).
}
\end{figure}

Given a network with two communities generated by the stochastic block model, we assign an initial state (red or blue) to each vertex.
We assume that $C_1$ is initially entirely red, and $C_2$ is entirely blue.
We refer to this state as a ``polarized'' initial condition.
This situation can occur, for instance, if the communities were previously unconnected, and each of them developed an internal consensus before new inter-community edges enabled interactions between communities.
(In Appendix~\ref{app:non-polarized}, we show numerical results for a non-polarized initial condition.)
We then update the states according to either the SLP, RLP, or CLP.
In the SLP, we randomly choose a ``sender'' from all vertices in the network.
The sender then exports its state to a randomly chosen neighbor.
In the RLP, the direction of the exchange is inverted: we pick a random ``recipient'' that adopts the state of a randomly chosen neighbor.
In the CLP, we first chose a random edge.
A randomly chosen vertex on the edge adopts the state of the other vertex on the same edge.
In Fig.~\ref{fig:netw_with_2_communities}, we illustrate the differences between the three processes.

In all three investigated processes, we modeled random drift  dynamics as a continuous-time Markov chain.
In other words, the time intervals between consecutive updates are independent and identically distributed exponential random numbers.
We set the time unit so that the mean update rate per vertex is $1$ in all investigated processes (SLP, RLP, and CLP).
Updates were repeated until the vertices reached a consensus.
Because we restricted our study to finite, connected networks, the occurrence of a consensus in finite time was guaranteed~\cite{serrano_conservation_2009}.

\section{Heterogeneous mean-field theory}
\label{sec:hetero_mf}

A continuous-time Markov chain is fully specified by the transition rates $Q(x, y)$ at which the process moves from any state $x$ to a new state ${y \neq x}$.
If we wanted to represent the state of every  vertex in a network with $N$ vertices faithfully, we would have to distinguish between $2^N$ different states of the network.
All Monte Carlo simulations presented in this study are based on this exact agent-based paradigm.
We complemented the simulations with a heterogeneous mean-field theory, in which we denote the state of the system as $(\rho_1, \rho_2)$ if a fraction $\rho_1$ of vertices in $C_1$ and a fraction $\rho_2$ of vertices in $C_2$ are red.

We can split the degree $k_i$ of a vertex in $C_i$ into two contributions: $k_i = k_{i, \text{int}} + k_{i, \text{ext}}$, where $k_{i, \text{int}}$ is the number of neighbors in the same community, and $k_{i, \text{ext}}$ is the number of neighbors in the opposite community.
In the stochastic block model, $k_{i, \text{int}}$ is a random variable with a binomial distribution $B(N_i - 1, P_i)$ for vertices in $C_i$.
We made a heterogeneous mean-field approximation by replacing $k_{i, \text{int}}$ with the mean of the distribution, $k_{i, \text{int}} \approx (N_i - 1) P_i$.
Similarly, we replaced $k_{i, \text{ext}}$ with $X / N_i$.

\begin{table*}
\caption{\label{tab:Q}
Transition rates from the state $(\rho_1, \rho_2)$ in the investigated random drift processes.
In these expressions, $E_i$ is the mean number of edges in $C_i$ (Eq.~\ref{eq:E_i}).
We only list transitions with positive rates.
}
\begin{ruledtabular}
  \begin{tabular*}{\textwidth}{l|lll}
    New state $(y, z)$
    & \multicolumn{3}{c}{Transition rate matrix element $Q[(\rho_1, \rho_2), (y, z)]$}\\
    & sender-limited
    & recipient-limited
    & connection-limited\\[3pt]
    \hline
    \noalign{\vskip 3pt}
    $\left(\rho_1 + \frac{1}{N_1}, \rho_2\right)$
    & $(1 - \rho_1) \left(
    \frac{N_1^3 P_1 \rho_1}{2 E_1 + X} + \frac{N_2 X \rho_2}{2 E_2 + X}
    \right)$
    & $N_1 (1-\rho_1) \cdot
      \frac{N_1^2 P_1 \rho_1 \, + \, X \rho_2}
      {2 E_1 \, + \, X}$
    & $N (1 - \rho_1) \cdot \frac{N_1^2 P_1 \rho_1 + X \rho_2}{2(E_1 + E_2 + X)}$\\[6pt]
    $\left(\rho_1 - \frac{1}{N_1}, \rho_2\right)$
    & $\rho_1 \left(
    \frac{N_1^3 P_1 (1 - \rho_1)}{2 E_1 + X} + \frac{N_2 X (1 - \rho_2)}{2 E_2 + X}
    \right)$
    & $N_1 \rho_1 \cdot
      \frac{N_1^2 P_1 (1 - \rho_1) \, + \, X (1 - \rho_2)}
      {2 E_1 \, + \, X}$
    & $N \rho_1 \cdot \frac{N_1^2 P_1 (1 - \rho_1) + X (1 - \rho_2)}{2(E_1 + E_2 + X)}$\\[6pt]
    $\left(\rho_1, \rho_2 + \frac{1}{N_2}\right)$
    & $(1 - \rho_2) \left(
      \frac{N_1 X \rho_1}{2 E_1 + X} + \frac{N_2^3 P_2 \rho_2}{2 E_2 + X} 
    \right)$
    & $N_2 (1-\rho_2) \cdot
      \frac{X \rho_1 \, + \, N_2^2 P_2 \rho_2}
      {2 E_2 \, + \, X}$
    & $N (1 - \rho_2) \cdot \frac{X \rho_1 + N_2^2 P_2 \rho_2}{2(E_1 + E_2 + X)}$\\[6pt]
    $\left(\rho_1, \rho_2 - \frac{1}{N_2}\right)$
    & $\rho_2 \left(
    \frac{N_1 X (1 - \rho_1)}{2 E_1 + X} +
    \frac{N_2^3 P_2 (1 - \rho_2)}{2 E_2 + X}
    \right)$
    & $N_2 \rho_2 \cdot
      \frac{X (1 - \rho_1) \, + \, N_2^2 P_2 (1 - \rho_2)}
      {2 E_2 \, + \, X}$
    & $N \rho_2 \cdot \frac{X (1 - \rho_1) + N_2^2 P_2 (1 - \rho_2)}{2(E_1 + E_2 + X)}$\\
  \end{tabular*}
 \end{ruledtabular}
\end{table*}

With these approximations, we can derive all the transition rates.
For example, let us consider the transition from $(\rho_1, \rho_2)$ to $(\rho_1 + 1/N_1, \rho_2)$.
The transition occurs when a vertex in $C_1$ changes from blue to red.
In the SLP, the transition probability can be written as a sum of two probabilities $\Pi_1 + \Pi_2$, where $\Pi_i$ is the probability that the sender is a red agent in $C_i$ who sends its state to a blue agent in $C_1$.
The first probability takes the following form:
\begin{equation}
  \Pi_1(\rho_1) = \frac{N_1 \rho_1}{N} \cdot  \frac{k_{1, \text{int}} \, N_1 (1 - \rho_1)}{k_1 (N_1 - 1)}\ ,
  \label{eq:P_1IP}
\end{equation}
where the first factor on the right-hand side, $N_1 \rho_1 / N$, is the probability of choosing a red sender in $C_1$.
The second factor is the expected proportion of blue neighbors in $C_1$, conditioned on the selection of a red agent in $C_1$ as the sender.
The second probability follows analogously:
\begin{align}
  \Pi_2(\rho_1, \rho_2)
  = \frac{N_2 \rho_2}{N} \cdot \frac{k_{2, \text{ext}} (1 - \rho_1)}{k_2}\ .
  \label{eq:P_2IP}
\end{align}
The transition rate is thus
\begin{align}
  & Q\left[(\rho_1, \rho_2), \left(\rho_1 + \frac{1}{N_1}, \rho_2\right)\right]
  \label{eq:Q_IP}\\
  & \qquad = N [\Pi_1(\rho_1) + \Pi_2(\rho_1, \rho_2)],\nonumber
\end{align}
where the factor $N$ accounts for the fact that there are, on average, $N$ updates per unit time.
After inserting Eqs.~\eqref{eq:P_1IP} and \eqref{eq:P_2IP} into Eq.~\eqref{eq:Q_IP}, we obtain the result in the top left corner of Table~\ref{tab:Q}, where we use the mean number of internal edges in $C_i$,
\begin{equation}
  E_i = \frac{1}{2} N_i (N_i - 1) P_i,
  \label{eq:E_i}
\end{equation}
as an auxiliary variable to shorten the expressions.
Based on similar arguments, we can calculate the other transition rates listed in Table~\ref{tab:Q}.
We adopted the convention that the diagonal elements of the transition rate matrix satisfy
\begin{equation}
  Q[(\rho_1, \rho_2), (\rho_1, \rho_2)] = -\sum_{\mathclap{\substack{(y, z)\\\neq (\rho_1, \rho_2)}}} Q[(\rho_1, \rho_2), (y, z)].
  \label{eq:Q_diag}
\end{equation}

\section{Probability of a red consensus}
\label{sec:prob_red_cons}

In this section, we calculate the probability $R(\rho_1, \rho_2)$ of reaching a red consensus from the initial state $(\rho_1, \rho_2)$. The choice of color is arbitrary; the labels ``red'' and ``blue'' are interchangeable.
$R$ is the martingale that satisfies $R(0, 0) = 0$, $R(1, 1) = 1$, and
\begin{equation}
  \sum_{y, z} Q[(\rho_1, \rho_2), (y, z)] R(y, z) = 0
  \label{eq:QR=0}
\end{equation}
for all $(\rho_1 , \rho_2) \notin \{(0, 0), (1, 1)\}$.
The solution for the investigated processes has the form
\begin{equation}
  R(\rho_1, \rho_2) = \frac{r_{12} \rho_1 + r_{21} \rho_2}{r_{12} + r_{21}}
  \label{eq:R}
\end{equation}
with
\begin{equation}
  r_{ij} =
  \begin{cases}
    N_i^2 (2 E_j + X) & \text{SLP,}\\
    2 E_i + X & \text{RLP,}\\
    N_i & \text{CLP.}
  \end{cases}
  \label{eq:r_i}
\end{equation}
It is possible to verify Eq.~\eqref{eq:R} and Eq.~\eqref{eq:r_i} by inserting them into Eq.~\eqref{eq:QR=0} together with the expressions for $Q[(\rho_1, \rho_2), (y, z)]$ in Table~\ref{tab:Q}.
We note that $r_{ij}$ for the SLP is not equal to $r_{ji}$ in the RLP; thus, there is no symmetry between the two processes.

Equation~\eqref{eq:r_i} shows that the probability of a red consensus in the SLP and RLP depends on the number of edges within and between communities.
In the CLP, however, $R$ is always the initial fraction of red vertices in the entire network, regardless of the details of the community structure.
Consequently, the initial majority state is always the likely consensus state in the CLP.
In the SLP and RLP, the initial majority is not necessarily the likely winner, as we can see from the following example.

\begin{figure}
    \centering
    \includegraphics[width=\columnwidth]{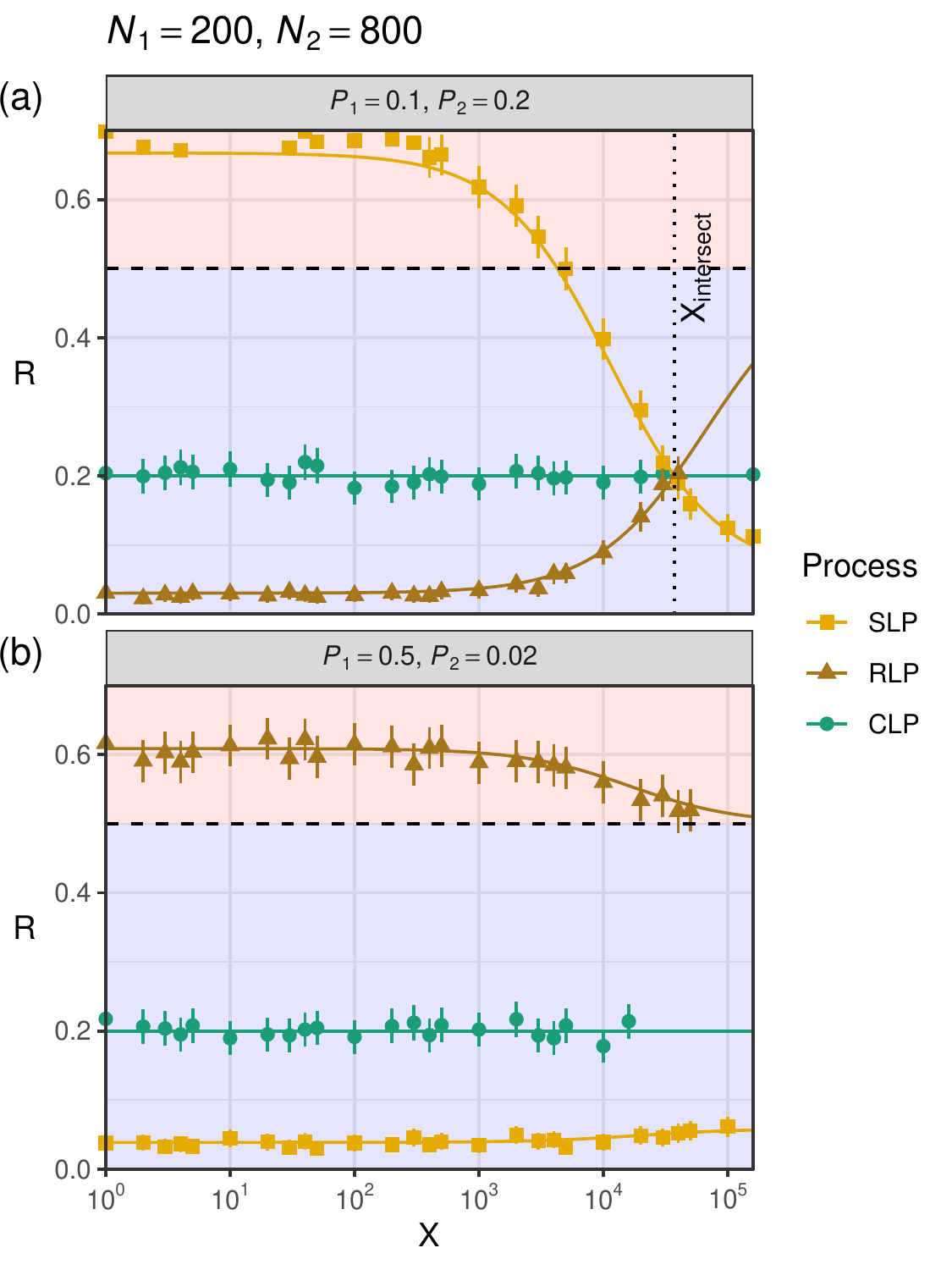}
    \caption{Probability $R$ of a red consensus as a function of the number of inter-community edges $X$ for two illustrative parameter sets.
    In all cases, the initial state is polarized: $(\rho_1, \rho_2) = (1, 0)$.
    Point symbols represent the means obtained from Monte Carlo simulations (error bars: 95\% confidence intervals).
    The curves are the theoretical predictions from Eqs.~\eqref{eq:R} and~\eqref{eq:r_i}.
    The background color (red vs.\ blue) indicates the likely winner. 
    The dotted line in panel (a) shows the inter-community connectivity for which the three processes have identical heterogeneous mean-field theories.
    For the parameters in panel (a), the smaller community $C_1$ has fewer internal edges than the larger community $C_2$.
    In panel (b), $C_1$ has more internal edges than $C_2$.}
    \label{fig:R}
\end{figure}

We assume a polarized initial state in which $\rho_1 = 1$ and $\rho_2 = 0$.
In general, if $R(\rho_1, \rho_2) > 1/2$, then the consensus opinion is more likely to be red than blue.
In the SLP, starting from a polarized state, this situation occurs if
\begin{equation}
  \frac{2 E_1 + X}{N_1^2} < \frac{2 E_2 + X}{N_2^2}\ .
  \label{eq:SLP_red_likely_winner}
\end{equation}
The quantity in the numerator, $2 E_i + X$, is the number of end points of the edges (also called ``edge stubs'') in $C_i$.
Equation~\eqref{eq:SLP_red_likely_winner} implies that even if $N_1 < N_2$ (i.e., red is initially in the minority), red is the likely final winner as long as the number of stubs in the red community is sufficiently small.
In Fig.~\ref{fig:R}, we confirmed the predictions of Eqs.\ \eqref{eq:R} and~\eqref{eq:r_i} using Monte Carlo simulations for an illustrative set of parameters.
In the SLP, the probability of a minority takeover from a polarized state decreases with $X$ if the minority community contains fewer edges than the majority (Fig.~\ref{fig:R}a).
However, even for maximally interconnected communities (i.e., $X = N_1 N_2$), the minority can be the likely winner; for example, if $N_1 = 100$, $E_1 = 0$, $N_2 = 150$, and $E_2 = 11\,175$ (in which case $C_2$ is a clique), then Eq.~\eqref{eq:r_i} implies that red wins with a probability of $R = 0.525$ despite initially having only 40\% of the votes.
This mean-field approximation agrees well with the Monte Carlo simulations ($R = 0.526$, 95\% confidence interval $[0.519, 0.533]$).

In the SLP, it is beneficial for the initial minority to have as few edge stubs as possible because, under these conditions, the minority is relatively rarely ``infected'' with the majority state and can spread its own state with high probability.
The opposite is true for the RLP.
If the RLP starts from $(\rho_1, \rho_2) = (1, 0)$, then $R$ is the fraction of edge stubs that are located in the red community $C_1$.
It follows that red is likely to win if $E_1 > E_2$ (Fig~\ref{fig:R}b).
It should be noted that this criterion does not depend on $X$.
If there are more edges between the communities, the difference between the success probabilities of the two states decreases.
However, the likely winner in the RLP is always the state that initially occupies the community with more internal edges.
Consequently, the likely winner can be the initial minority  (e.g., red if $N_1 < N_2$ and $E_1 > E_2$, see the RLP curve in Fig.~\ref{fig:R}b).

In general, the SLP, RLP, and CLP have different values of $R$.
An exception is the case in which the ratio of the number of vertices in the two communities is equal to the ratio of the edge stubs,
\begin{align}
  \frac{N_1}{N_2} = \frac{2 E_1 + X}{2 E_2 + X}\ .
  \label{eq:equal_processes}
\end{align}
In this case, it follows from Eqs.~\eqref{eq:R} and~\eqref{eq:r_i} that $R$ equals the initial fraction of red vertices in all three processes.
An intuitive way to understand this feature is to note that the left-hand side of Eq.~\eqref{eq:equal_processes} describes the SLP's (RLP's) odds of choosing a sender (recipient) in $C_1$ vs.\ $C_2$ during one update.
The right-hand side represents the odds of choosing the SLP's recipient (RLP's sender) from the vertices in $C_1$ vs.\ $C_2$ in the heterogeneous mean-field approximation.
In the CLP, the right-hand side is equal to the odds of being a sender and the odds of being a recipient.
If Eq.~\eqref{eq:equal_processes} is satisfied, then all odds mentioned above are equal; thus, the processes are identical under the heterogeneous mean-field assumption.
In Fig.~\ref{fig:R}(a), we confirm this prediction with Monte Carlo simulations for an illustrative set of parameters: $R$ is the same for the SLP, RLP, and CLP if $X$ has the special value 
\begin{equation}
X_\text{intersect} = \frac{N_1 N_2 [(N_2 - 1) P_2 - (N_1 - 1) P_1]} {N_2 - N_1}\ .
\label{eq:X_intersect}
\end{equation}
If Eq.~\eqref{eq:X_intersect} formally predicts $X_\text{intersect} < 0$ (e.g., for the parameters chosen in Fig.~\ref{fig:R}b), no intersection exists.

\section{Mean consensus time}
\label{sec:mean_cons_time}

The transition rates listed in Table~\ref{tab:Q} determine the mean time $T(\rho_1, \rho_2)$ until a consensus is reached from the initial state $(\rho_1, \rho_2)$.
The equations that determine $T$ are $T(0, 0) = T(1, 1) = 0$ and
\begin{equation}
  \sum_{y, z} Q[(\rho_1, \rho_2), (y, z)] \, T(y, z) = -1
  \label{eq:T_general}
\end{equation}
for all $(\rho_1, \rho_2) \notin \{(0, 0), (1, 1)\}$.
When we insert the transition rates from Table~\ref{tab:Q} and Eq.~\eqref{eq:Q_diag} into Eq.~\eqref{eq:T_general}, we can obtain recurrence relations for the values of $T$.
Using a diffusion approximation~\cite{ewens_mathematical_2004}, we can convert the recurrence relations into the more familiar form of a partial differential equation.
That is, we assume $N_1, N_2 \gg 1$ and take the continuum limit of Eq.~\eqref{eq:T_general}.
For the SLP, this procedure leads to the following partial differential equation:
\begin{align}
  & \left(
    \frac{
      N_1 P_1 \, \rho_1 (1 - \rho_1)
    }{
      2 E_1 + X
    } +
    \frac{
      N_2 X (\rho_1 + \rho_2 - 2 \rho_1 \rho_2)
    }{
      2 N_1^2 (2 E_2 + X)
    }
  \right)
  \frac{\partial^2 T}{\partial \rho_1^2}
  \nonumber\\
  & + \left(
    \frac{
      N_2 P_2 \, \rho_2 (1 - \rho_2)
    }{
      2 E_2 + X
    } +
    \frac{
      N_1 X (\rho_1 + \rho_2 - 2 \rho_1 \rho_2)
    }{
      2 N_2^2 (2 E_1 + X)
    }
  \right)
  \frac{\partial^2 T}{\partial \rho_2^2}
  \nonumber\\
  & + \frac{
    N_2 X (\rho_2 - \rho_1)
  }{
    N_1 (2 E_2 + X)
  } \frac{\partial T}{\partial \rho_1} +
  \frac{
    N_1 X (\rho_1 - \rho_2)
  }{
    N_2 (2 E_1 + X)
  } \frac{\partial T}{\partial \rho_2}
  \nonumber\\
  & = - 1.
  \label{eq:T_IP_general_PDE}
\end{align}

We now assume that the communities are so sparsely connected to each other that $X \ll \min(N_1, N_2)$.
We can then drop all terms from Eq.~\eqref{eq:T_IP_general_PDE} that are $O\left(N_1^{-2}\right)$ or $O\left(N_2^{-2}\right)$,
\begin{align}
& \frac{\rho_1 (1-\rho_1)}{N_1} \frac{\partial^2 T}{\partial \rho_1^2} +
\frac{\rho_2 (1-\rho_2)}{N_2} \frac{\partial^2 T}{\partial \rho_2^2}
\label{eq:T_SLP_for_small_X}\\
& +\frac{X (\rho_2 - \rho_1)}{N_1 N_2 P_2} \frac{\partial T}{\partial \rho_1} + \frac{X (\rho_1 - \rho_2)}{N_1 N_2 P_1} \frac{\partial T}{\partial \rho_2} = -1.
\nonumber
\end{align}
For the RLP and CLP, we obtain similar second-order partial differential equations (see Appendix~\ref{app:small_X}).

\begin{figure}
    \centering
    \includegraphics[width=\columnwidth]{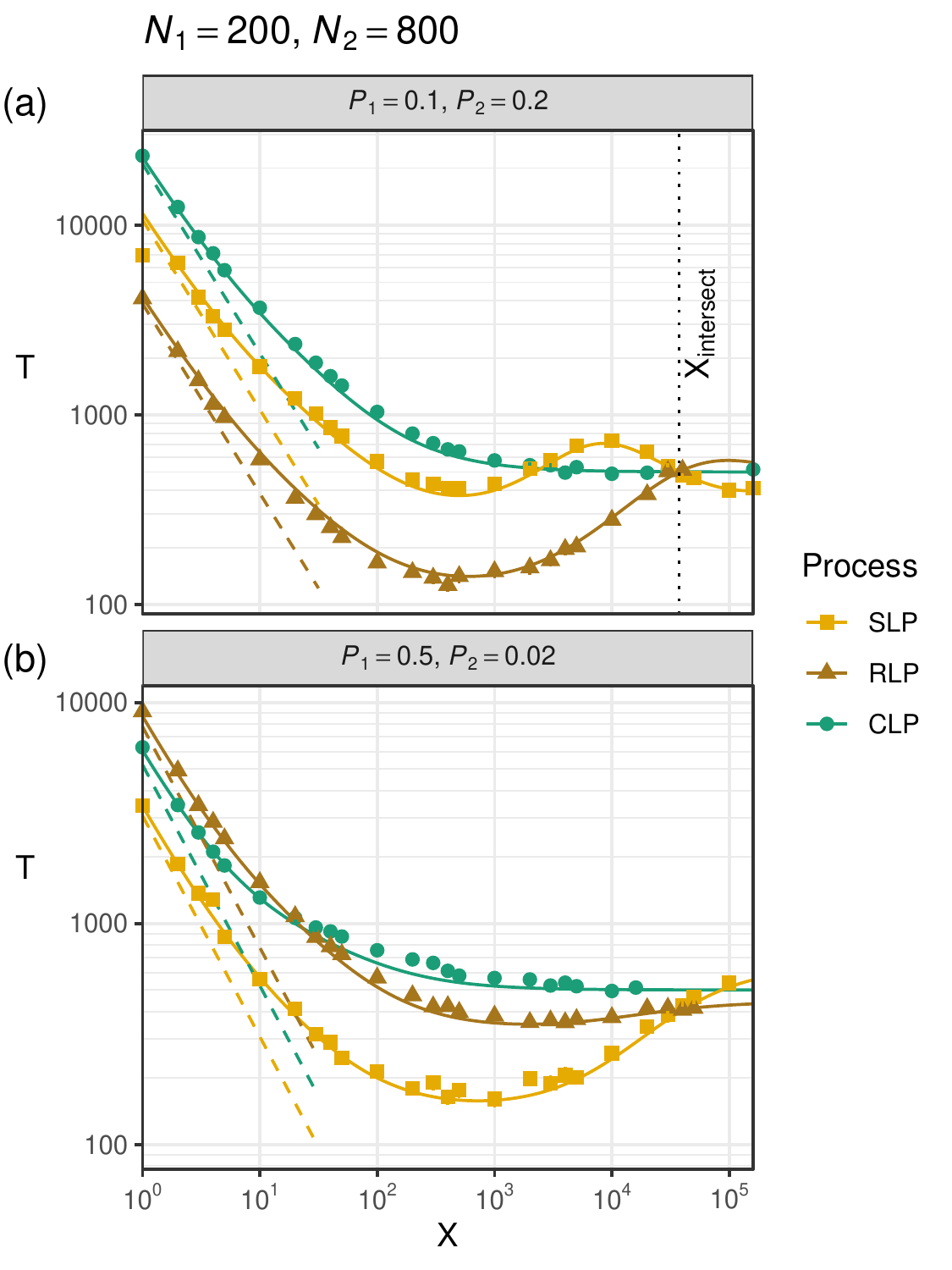}
    \caption{Mean consensus time $T$ as a function of the number of inter-community edges $X$ for two illustrative parameter sets.
    In all cases, the initial state is polarized: $(\rho_1, \rho_2) = (1, 0)$.
    Point symbols represent Monte Carlo simulations. Error bars are invisible because the 95\% confidence intervals are smaller than the symbol sizes.
    The curves represent numerical solutions of the diffusion approximation (Eqs.~\ref{eq:T_IP_general_PDE} and~\ref{eq:app_diff_approx}), obtained using the method described in Appendix~\ref{app:arbitrary_X}.
    The dashed lines show the approximations of Eq.~\eqref{eq:T_sparse_10} for $X \ll \min(N_1, N_2)$.
    The dotted line in panel (a) shows the inter-community connectivity for which the three processes have identical heterogeneous mean-field theories (Eq.~\ref{eq:X_intersect}).}.
    \label{fig:T}
\end{figure}

We are not aware of a closed-form solution of Eq.~\eqref{eq:T_SLP_for_small_X} or the corresponding equations for the RLP and CLP.
However, we can obtain a good approximation of $T$ using a two-dimensional power series.
We call this approximation $T_\text{sparse}$ to indicate that this approximation is valid only if the communities are sparsely connected to each other,
\begin{align}
  & T_\text{sparse}(\rho_1, \rho_2)
  \label{eq:T_sparse}\\
  & \qquad = \sum_{i=0}^2 \sum_{j=0}^2 c_{ij} \left(\rho_1 - \frac{1}{2}\right)^i \left(\rho_2 - \frac{1}{2}\right)^j.
  \nonumber
\end{align}
We expand the right-hand side only up to quadratic terms because this approximation is sufficiently accurate if $X$ is small.
The value of $T$ remains unchanged if we swap the labels of the states (i.e., red and blue); thus, $T$ must satisfy $T(\rho_1, \rho_2) = T(1 - \rho_1, 1 - \rho_2)$.
Consequently, the coefficients $c_{ij}$ in Eq.~\eqref{eq:T_sparse} must be zero if either $i$ is odd and $j$ is even, or vice versa.
The remaining coefficients can be determined using Eq.~\eqref{eq:T_SLP_for_small_X} and the boundary conditions.
In Appendix~\ref{app:small_X}, we show how to cast the conditions on the coefficients into a system of linear equations for the five unknowns $c_{00}$, $c_{02}$, $c_{11}$, $c_{20}$, and $c_{22}$.
From the leading-order behavior of $c_{ij}$, we can infer the consensus time for the polarized initial conditions in the limit ${\frac{X}{\min(N_1, N_2)} \to 0}$,
\begin{align}
  T_\text{sparse}(1, 0) = 
  \frac{1}{X} \cdot
  \begin{cases}
    \frac{N_1 N_2 P_1 P_2}{P_1 + P_2} & \text{SLP,}\\[6pt]
    \frac{2 E_1 E_2}{E_1 + E_2} & \text{RLP,}\\[6pt]
    \frac{2 N_1 N_2 (E_1 + E_2)} {N^2} & \text{CLP.}
  \end{cases}
  \label{eq:T_sparse_10}
\end{align}
The dashed lines in Fig.~\ref{fig:T} indicate these limits.

We can deduce from Eq.~\eqref{eq:T_sparse_10} that the mean consensus time in the SLP for a small $X$ is always shorter than in the CLP.
The mean consensus time for the RLP can range from being shorter than for the SLP (if the community with more vertices has a higher average within-community degree) to being longer than for the CLP (if $P_1 / P_2$ is between $N_2 / N_1$ and $N_2^3 / N_1^3$).
The two illustrative parameter sets shown in Fig.~\ref{fig:T} exhibit both cases.

The points in Fig.~\ref{fig:T} indicate the results from the Monte Carlo simulations.
The curves in Fig.~\ref{fig:T} are numerical solutions of the diffusion approximations for the SLP, RLP, and CLP, calculated using the method described in Appendix~\ref{app:arbitrary_X}.
As $X$ increases, the leading-order approximation in Eq.~\eqref{eq:T_sparse_10} becomes inaccurate, and the rankings of the processes in terms of $T$ change.
However, the agreement with the heterogeneous mean-field theory (curves in Fig.~\ref{fig:T}) is excellent.
For the special inter-community connectivity $X_\text{intersect}$, derived in Eq.~\eqref{eq:X_intersect}, the mean consensus times for all processes are equal.
Besides this special case, there are no simple rules that govern which process is fastest or slowest for a given $X$.

While $T$ is a monotonically decreasing function of $X$ for the CLP, there can be local minima and maxima for the SLP and RLP.
Thus, it is possible to increase the mean consensus time by increasing the inter-community connectivity.
At first glance, this effect is counterintuitive because one might expect that more inter-community edges, which necessarily speed up communication between the cliques, would always lead, on average, to a faster consensus.
The resolution to this apparent paradox lies in the fact that a larger $X$ can give the initial minority state more options to displace the majority, which can slow down consensus formation.
For the special case of the RLP on two cliques, this counterintuitive effect has already been observed by Gastner and Ishida~\cite{gastner_voter_2019}.

So far, we have focused on polarized initial states.
In Appendix~\ref{app:non-polarized}, we present results for a non-polarized initial state with $\rho_1 = \rho_2 = 1/2$.
As for the polarized case, we found that $T$ is a monotonically decreasing function of $X$ for the CLP but not for the SLP and RLP (Fig.~\ref{fig:T_unpolarized}).
If $X$ is small, we also observed the same rankings of the processes in terms of $T$.
For larger values of $X$, however, the rankings generally depend on the initial state.

\section{\label{sec:discussion}Discussion}

We compared the effect of three limitations on spreading in networks that had a specific community structure: each network consisted of two subgraphs (communities) with arbitrary sizes and connected by $X$ edges.
The parameter $X$ varied from $1$, representing a single ``bottleneck'' transmission channel (or ecological corridor) between the communities, up to $X=N_1 N_2$, in which case every possible inter-community connection was present.

\subsection{Two communities divided by a barrier}

The case of a small $X$ is particularly interesting from a biological viewpoint.
In an ecological interpretation, the model can be viewed as a two-species metacommunity model, in which the vertices represent discrete habitat sites.
A barrier (e.g., mountain range or river) divides the area into two regions.
$X$ represents the penetrability of the barrier.
For example, the barrier can be transgressed via ecological corridors.

Notably, when the barrier is hardly permeable (i.e., $X$ is very small), the time of winning is inversely related to $X$ (Eq.~\ref{eq:T_sparse_10}).
For example, when the two sides of a barrier are connected by twice as many corridors, the time of coexistence is halved.
This rule is important when we prefer coexistence, to maintain species diversity, or when we wish to eliminate one of the species.
In real life, an additional corridor can arise naturally (e.g., by a rock slide on a mountain), or can be man-made, either voluntarily or involuntarily.
A typical example of the latter is when building a road, railway line, or canal promotes the spread of species (e.g., weeds) from one geographic region into another.
Our results suggest that having two corridors instead of one does not considerably influence which species wins (Fig.~2) but dramatically influences the time required for winning (Fig.~3). 

Although our model contains serious simplifications, it reflects two important, realistic features of real-life metacommunities.
First, local dynamics are integrated into regional dynamics~\cite{economo_biodiversity_2011, thompson_process-based_2020}.
Second, the opportunity for spreading of a species depends on the initial position within the network (see, for example, a review on river networks in~\cite{borthagaray_chapter_2015}).
The capacity of the voter model for studying metacommunities has only rarely been used (e.g.,~\cite{chave_spatial_2001}); to our knowledge, our model is the first to introduce a habitat structure that has the topology of a modular network (local patches with a regional barrier).

The effect of a modular habitat structure is similarly important in evolutionary models, in which the colors (red and blue) represent genotypes within a species.
Geographic barriers have often been mentioned as major drivers of evolution, including speciation~\cite{coyne_speciation_2004}.
Therefore, studying the penetrability of barriers, compared to spreading within each side, is of primary importance.
We propose that network theory can provide considerable help in this regard.

Our model is also applicable to social networks with exogenous community structures.
Barriers in social networks can be along religious divisions (e.g., Catholics vs.\ Protestants in Northern Ireland), language barriers (e.g., between Dutch and French speakers in Belgium), or ethnic conflicts (e.g., between Greek and Turkish Cypriots).
The likelihood of inter-community links can be increased, for example, by working in the same place, visiting common places of entertainment, or living in the same neighborhood. 
The present results suggest that, when the number of links is originally low, adding a few new links can significantly decrease the time required for consensus formation.

\subsection{Validity of the heterogeneous mean-field theory}

The heterogeneous mean-field approximation in Sec.~\ref{sec:hetero_mf} assumes perfect mixing (in ecological terms, no dispersal limitation) within each side of the barrier, whereas our simulated networks represent the exact configuration of the links (ecological corridors) between the sites.
Interestingly, the heterogeneous mean-field approximation agrees well with the results of the explicit (i.e., agent-based) model.
Because the heterogeneous mean-field approximation does not contain detailed information regarding the network structure, its effectiveness in predicting the outcome could be utilized in ecological fieldwork and observational studies of social dynamics: it is easier to estimate the connection probability within and between groups of vertices than to map the links exactly.

At small $X$, the agreement between the agent-based model and the heterogeneous mean-field theory can be explained by the fact that winning within a single side (within a community) is a relatively fast process because of the higher intra-community connectivity.
The main bottleneck for an invader is to find a way from one community to another.
Therefore, all sites within the same side of the barrier can be considered almost identical in terms of sending and receiving the species.

The validity of the heterogeneous mean-field theory is not limited to a small $X$.
If $X$ is large, the proportion of red vertices $\rho_1$ and $\rho_2$ quickly equalizes between communities $1$ and $2$.
In this case, the heterogeneous mean-field theory approximates the random drifts in $\rho_1$ and $\rho_2$ as being coupled by the constraint $\rho_1 = \rho_2$~\cite{gastner_impact_2019, gastner_voter_2019}; thus, the vertices in the same community can again be treated as almost identical.
Figure~\ref{fig:R} shows that the theory correctly predicts that, in the SLP, the initial minority can be the likely consensus at small $X$ but is unlikely to win at large $X$, even if all other parameters ($P_1$, $P_2$, and the initial distribution of red and blue vertices) are held constant.
The heterogeneous mean-field approximation also correctly captures the sensitive dependence of the consensus time $T$  on the process (SLP, RLP, and CLP) and the network parameters ($P_1$, $P_2$, $N_1$, and $N_2$) for the full range of inter-community connectivity (i.e., from $X=1$ to $X=N_1 N_2$; see Fig.~\ref{fig:T}).

As $X$ increases, the network topology changes from a modular structure sensu stricto (i.e., two clusters divided by a barrier) over a core-periphery structure to an anti-community structure, where both communities are sparsely connected internally and highly connected to each other.
Ecological habitat networks often form a core-periphery structure toward the edge of the geographic range~\cite{safriel_core_1994}, as the suitable area tends to become fragmented~\cite{gastner_transition_2009}. 
However, these cases require more complex models; the network's parameters are unlikely to be homogeneous from the core to the periphery because of a change in the environment across space.
More typical examples of core-periphery networks are available in social sciences.
For example, networks of professional relationships between scientists have been reported to possess a core-periphery structure~\cite{brieger_career_1976, freeman_impact_1984}.
Networks of romantic relationships typically exhibit anti-community structures because most edges are between agents of different genders~\cite{bearman_chains_2004}.
Our results underline the importance of knowing the limiting process (SLP, RLP, or CLP) regardless of whether the network has a modular, core-periphery, or anti-community structure.

\subsection{Effect of sender, recipient, and channel limitations}

The importance of the update rule has been emphasized in several studies on general networks (e.g.,~\cite{sood_voter_2008, maciejewski_reproductive_2014, masuda_directionality_2009, moretti_heterogenous_2012}), and various update rules have appeared under unrelated names. 
The nomenclature suggested here (SLP, CLP, and RLP) expresses that there are different limiting factors in the process of copying the state of a vertex to one of its neighbors.
The spread can be limited by the sender (as in our SLP), the connection (CLP), or the recipient (RLP).
In our study, only one type of limitation was present in each case.
To our knowledge, earlier studies have also assumed only one type of limitation in each process.
For example, Sood et al.~\cite{sood_voter_2008} studied the voter model (equivalent to our RLP) and two processes that they termed the ``invasion process'' (equivalent to our SLP) and ``link dynamics'' (our CLP).
Castellano~\cite{castellano_effect_2005} called the SLP a ``reverse voter model'', whereas
Macijewski~\cite{maciejewski_reproductive_2014} discussed the SLP and RLP under the names birth--death and death--birth Moran processes, respectively.

In an ecological and evolutionary context, the SLP, CLP, and RLP represent different types of limitation on the species/genotypes.
In the SLP, the production of dispersers (e.g., seeds in plants) is limited.
The CLP corresponds to another type of dispersal limitation; here, the movement of the species between habitat patches is limited.
This situation occurs, for instance, when the distance between the patches is high or the penetrability of the terrain is low, relative to the organisms' ability to move.
For example, if the landing distance of seeds from the parent plant is relatively short compared to the distance between suitable habitat patches, the CLP is applicable.
Finally, the RLP represents establishment limitation.
In the case of a plant species, establishment limitation typically occurs when germination in the new site is limited, or the survival probability of the seedlings is low.

Several studies on metacommunity ecology and landscape ecology have shown that these factors are important for the maintenance of biodiversity in landscapes with multiple habitat patches~\cite{ehrlen_dispersal_2000}.
For example, the pace of succession in plant communities can be significantly reduced by any of the aforementioned limiting factors~\cite{oster_dispersal_2009}. 

In this study, we compared the relative importance of the three types of limiting factors. 
Let us consider the case in which the initial condition is polarized: region 1 is inhabited solely by the ``red'' species ($\rho_1=1$), whereas region 2 is occupied by the ``blue'' species ($\rho_2=0$). 
Without loss of generality, we assume that $N_1<N_2$, that is, the red species is less frequent than the blue one (it is a ``minority'').
For example, the red color may represent a species that has newly arrived as a potential invader and is attempting to cross a barrier for the first time.
Alternatively, red and blue can represent two vicariant species that have lived stably on different sides of the barrier~\cite{humphries_vicariance_2017}.
The process starts when the originally impermeable barrier becomes permeable ($X>0$), for example, due to a climate change.
Building roads or bridges can also make a barrier permeable, depending on the species.
Creating artificial ecological corridors is a means of increasing permeability for species that are worthy of protection.
Even a single link ($X=1$) initiates a diffusion of species across the barrier.
The endpoint is reached when one of the species outcompetes the other on both sides of the barrier.
If ``red'' is an alien invader, then low $R$ (probability of red fixation, see Sec.~\ref{sec:prob_red_cons}) is preferred from the perspective of nature conservation.
Conversely, if ``red'' is a rare species that is worthy of protection, a high $R$ may be desirable.
Our results suggest that there are two scenarios in which invasion across the barrier by the initial minority is likely (see the red region in Fig.~\ref{fig:R}): in the SLP, when the internal connectivity on the ``red'' side of the barrier is low compared to the ``blue'' side, or in the RLP, when this connectivity is high.
This finding agrees with earlier results in the opinion dynamics literature: it depends on the update rule whether low-degree or high-degree vertices are more likely to spread their opinions~\cite{sood_voter_2008, moretti_heterogenous_2012, voorhees_birthdeath_2013, tan_characterizing_2014, iwamasa_networks_2014, maciejewski_reproductive_2014}.
Notably, invasion by the minority can be successful not only at low $X$ but also in a broad range of $X$, depending on $P_1$ and $P_2$ (Fig.~\ref{fig:R}).
In general, starting from good network positions, the minority can win against the majority, even though it does not enjoy any local advantage (i.e., the competition is neutral).

If spreading is limited because crossing the corridors is difficult (i.e., in the CLP), then the final outcome of competition does not depend on the number of corridors ($X$) and only the time of winning does.
Therefore, creating more ecological corridors across a barrier or closing some existing corridors can influence the pace of invasion, but not the probability of winning, at least against an alternative species in neutral competition.
In the case of the CLP, the network positions do not matter either: it is merely the initial abundance of the two species that determines $R$. In other words, we observe a mass effect.

\subsection{Suggestions for future work}

The aforementioned results hold true only for neutral competition between two species/genotypes.
To gain a more general view, the model should be extended to non-neutral competition and to more than two species/genotypes.
Non-neutrality (i.e., selection) has been introduced into several network models of evolutionary processes~\cite{lieberman_evolutionary_2005, antal_evolutionary_2006, maciejewski_reproductive_2014}.

In the present model, a considerable simplification is that each vertex can be inhabited only by a single species (red or blue) at a time.
This assumption is plausible in two cases.
In the first case, the spatial resolution is so fine that each site can contain only a single individual (that belongs to the red or blue species).
In the second case, a time scale separation can be made, assuming that the time needed for competitive exclusion within a site is much shorter than the waiting time for the arrival of a new colonizer (i.e., the system is strongly dispersal-limited).
Future extensions of the model with non-binary discrete or continuous states are worthy of investigation.
The literature on social influence has highlighted substantial differences between opinion dynamics on binary, continuous, and nominal scales~\cite{flache_models_2017}.

Another interesting task for future research is to investigate two or three types of limitations on spreading acting simultaneously.
Their relative importance can be thought of as points in a multi-dimensional parameter space, in which we have so far only investigated special cases when two out of three processes (sending, receiving, and transmitting) occur at infinite rates.

We acknowledge that the network model presented in this study is not spatially explicit.
However, it would be straightforward to replace the non-spatial stochastic block model with a model for modular spatial networks (e.g., the model proposed by Gross et al.~\cite{gross_two_2020}).
Equation-based results are difficult to obtain for spatially explicit models; however, Monte Carlo simulations are certainly possible.

In summary, we studied three different agent-based models of neutral competition using equation-based and numerical techniques.
The update rules of the models assumed that either the senders, recipients, or channels of transmission were the bottlenecks in the spread of the states from one vertex to another.
While we acknowledge that our update rules and network models are highly simplified compared to real-world applications, we believe that our results provide a basis for future studies of agent-based competition in modular networks.

\begin{acknowledgments}
This work was supported by the Singapore Ministry of Education (MOE) and Yale-NUS College (through grant number R-607-263-043-121) and the National Science Foundation of Hungary (HU NKFI FK K124438).
We would like to thank Editage (www.editage.com) for English language editing.
\end{acknowledgments}

\appendix

\section{\label{app:non-polarized}Effect of a non-polarized initial condition}

All numerical results in the main text were obtained using polarized initial states.
In this appendix, we compare those results with the outcomes of Monte Carlo simulations for an illustrative non-polarized initial state: $\rho_1 = \rho_2 =  1/2$.
We randomly assign an initial color (red or blue) to each vertex with probability $1/2$.
In this case, Eq.~\eqref{eq:R} predicts $R=1/2$ for all $X$ and for all three processes (SLP, RLP, and CLP).
Thus, if both communities initially have an equal number of vertices of both colors, then both colors are equally likely to win regardless of the number of inter-community links and the process.
We confirmed this prediction with Monte Carlo simulations.

\begin{figure}
    \centering
    \includegraphics[width=\columnwidth]{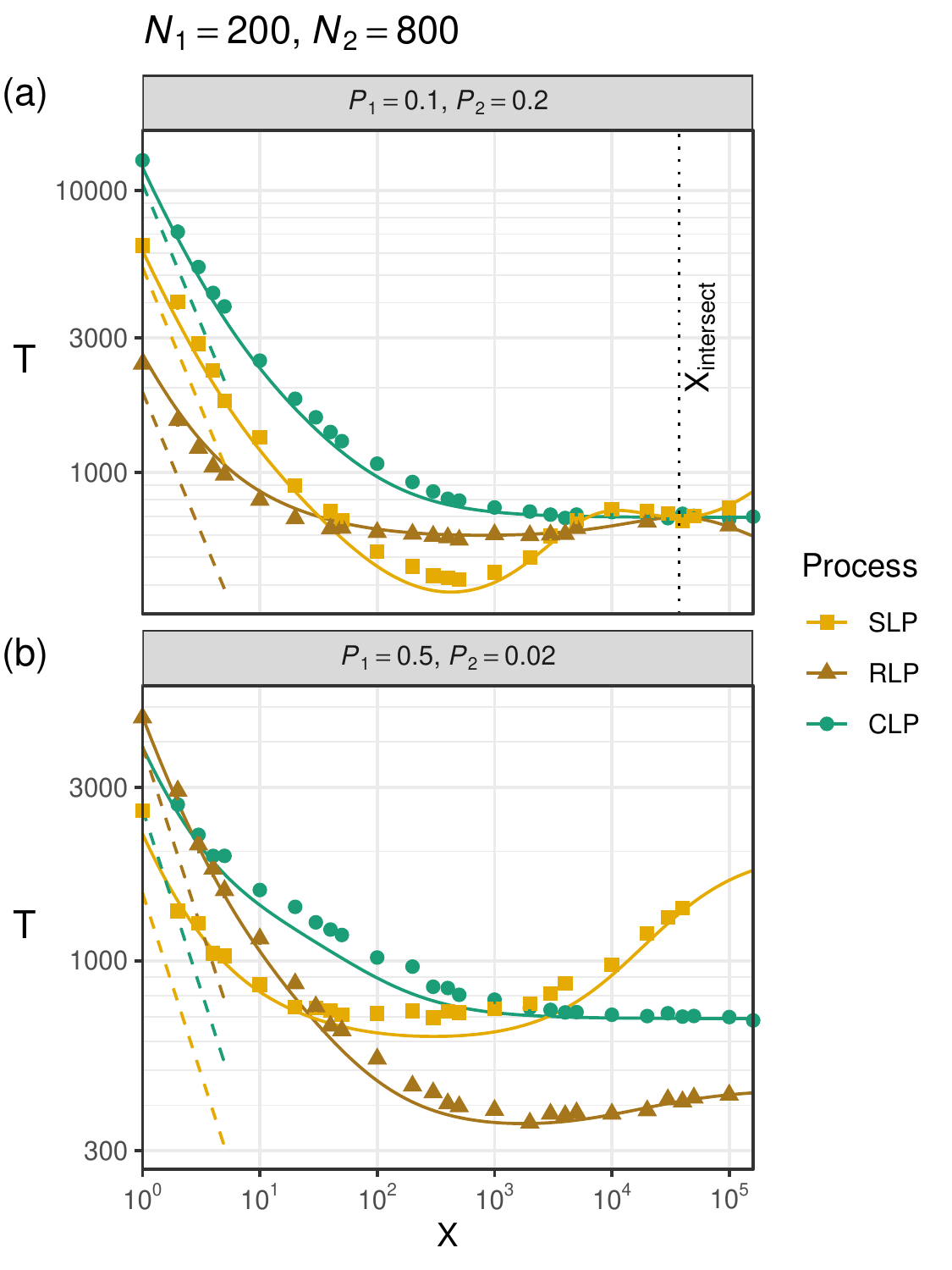}
    \caption{Mean consensus time $T$ as a function of the number of inter-community edges $X$ for a non-polarized initial state: $\rho_1 = \rho_2 = 0.5$.
    The parameters $N_1$, $N_2$, $P_1$ and $P_2$ are the same as in Fig.~\ref{fig:T}.
    Point symbols represent Monte Carlo simulations. 
    The curves represent numerical solutions of the diffusion approximation (Eqs.~\ref{eq:T_IP_general_PDE} and~\ref{eq:app_diff_approx}), obtained using the method described in Appendix~\ref{app:arbitrary_X}.
    The dashed lines show the approximation of Eq.~\eqref{eq:T_sparse_unpolarized_from_c11} for $X \ll \min(N_1, N_2)$.
    \label{fig:T_unpolarized}}
\end{figure}

Unlike $R$, the mean consensus time $T$ depends on $X$ and the type of process (Fig.~\ref{fig:T_unpolarized}).
In Appendix~\ref{app:small_X}, we present an analytic technique that can approximate $T$ as a function of $X$.
This approximation is applicable for all initial conditions.
In Fig.~\ref{fig:T_unpolarized}, we represent the approximation by the solid curves.
The approximation is in good agreement with Monte Carlo simulations (represented by point symbols in Fig.~\ref{fig:T_unpolarized}).
In the limit of small $X$, it follows from Eqs.~\eqref{eq:T_sparse_10_from_c11} and~\eqref{eq:T_sparse_unpolarized_from_c11} that the mean consensus time in the non-polarized case is half of that in the polarized case; thus, the rankings of the processes in this limit are the same for both types of initial conditions (polarized and non-polarized with $\rho_1 = \rho_2 = 1/2$).
For larger values of $X$, however, there are no simple rules that would relate the mean consensus time to the initial conditions.

\section{\label{app:small_X}Mean consensus time for sparse inter-community connectivity}

The diffusion approximation for the three processes under investigation (SLP, RLP, and CLP) has the following general form:
\begin{align}
& f_1(\rho_1, \rho_2)\, \frac{\partial^2 T}{\partial \rho_1^2} +
f_2(\rho_1, \rho_2)\, \frac{\partial^2 T}{\partial \rho_2^2}
\label{eq:app_diff_approx}\\
& + g_1(\rho_1, \rho_2)\, \frac{\partial T}{\partial \rho_1} + g_2(\rho_1, \rho_2)\, \frac{\partial T}{\partial \rho_2} = -1\nonumber
\end{align}
with functions $f_i(\rho_1, \rho_2)$ and $g_i(\rho_1, \rho_2)$ which differ between the processes.
These functions are listed in Table~\ref{tab:app_diff_approx}.
If $X=O[\min(N_1, N_2)]$, the leading-order terms of Eq.~\eqref{eq:app_diff_approx} are
\begin{align}
  & H_1 \, \rho_1 (1 - \rho_1) \frac{\partial^2 T}{\partial \rho_1^2} + H_2 \, \rho_2 (1 - \rho_2) \frac{\partial^2 T}{\partial \rho_2^2}
  \label{eq:T_sparse_general}\\
  & + I_1 \, X (\rho_2 - \rho_1) \frac{\partial T}{\partial \rho_1} + I_2 \, X (\rho_1 - \rho_2) \frac{\partial T}{\partial \rho_2} = -1,
  \nonumber
\end{align}
where the parameters $H_i$ and $I_i$ are dependent on the process.
Their values are listed in Table~\ref{tab:app_diff_approx}.

\begin{table*}
\caption{\label{tab:app_diff_approx}
Functions and parameters that appear in the diffusion approximation (Eqs.~\ref{eq:app_diff_approx} and~\ref{eq:T_sparse_general}).
}
\begin{ruledtabular}
  \begin{tabular*}{\textwidth}{l|lll}
    Function
    & sender-limited
    & recipient-limited
    & connection-limited\\[3pt]
    \hline
    \noalign{\vskip 3pt}
    $f_1(\rho_1, \rho_2)$
    & $\frac{N_1 P_1 \, \rho_1 (1 - \rho_1)}{2 E_1 + X} + \frac{N_2 X (\rho_1 + \rho_2 - 2 \rho_1 \rho_2)}{2 N_1^2 (2 E_2 + X)}$
    & $\frac{2 N_1^2 \, P_1 \, \rho_1 (1 - \rho_1) + X (\rho_1 + \rho_2 - 2 \rho_1 \rho_2)}{2 N_1 (2 E_1 + X)}$
    & $\frac{N P_1 \, \rho_1 (1 - \rho_1)}{2 (E_1 + E_2 + X)} + \frac{N X (\rho_1 + \rho_2 - 2 \rho_1 \rho_2)}{4 N_1^2 (E_1 + E_2 + X)}$\\[6pt]
    $f_2(\rho_1, \rho_2)$
    & $\frac{N_2 P_2 \, \rho_2 (1 - \rho_2)}{2 E_2 + X} + \frac{N_1 X (\rho_1 + \rho_2 - 2 \rho_1 \rho_2)}{2 N_2^2 (2 E_1 + X)}$
    & $\frac{2 N_2^2 \, P_2 \, \rho_2 (1 - \rho_2) + X (\rho_1 + \rho_2 - 2 \rho_1 \rho_2)}{2 N_2 (2 E_2 + X)}$
    & $\frac{N P_2 \rho_2 (1-\rho_2)}{2 (E_1 + E_2 + X)} + \frac{N X (\rho_1 + \rho_2 - 2 \rho_1 \rho_2)}{4 N_2^2 (E_1 + E_2 + X)}$\\[6pt]
    $g_1(\rho_1, \rho_2)$
    & $\frac{N_2 X (\rho_2 - \rho_1)}{N_1 (2 E_2 + X)}$
    & $\frac{X (\rho_2 - \rho_1)}{2 E_1 + X}$
    & $\frac{N X (\rho_2 - \rho_1)}{2 N_1 (E_1 + E_2 + X)}$\\[6pt]
    $g_2(\rho_1, \rho_2)$
    & $\frac{N_1 X (\rho_1 - \rho_2)}{N_2 (2 E_1 + X)}$
    & $\frac{X (\rho_1 - \rho_2)}{2 E_2 + X}$
    & $\frac{N X (\rho_1 - \rho_2)}{2 N_2 (E_1 + E_2 + X)}$\\[6pt]
    $H_1$
    & $N_1^{-1}$
    & $N_1^{-1}$
    & $\frac{N P_1}{N_1^2 P_1 + N_2^2 P_2}$\\[6pt]
    $H_2$
    & $N_2^{-1}$
    & $N_2^{-1}$
    & $\frac{N P_2}{N_1^2 P_1 + N_2^2 P_2}$\\[6pt]
    $I_1$
    & $(N_1 N_2 P_1)^{-1}$
    & $(N_1^2 P_1)^{-1}$
    & $\frac{N}{N_1 (N_1^2 P_1 + N_2^2 P_2)}$\\[6pt]
    $I_2$
    & $(N_1 N_2 P_2)^{-1}$
    & $(N_2^2 P_2)^{-1}$
    & $\frac{N}{N_2 (N_1^2 P_1 + N_2^2 P_2)}$
  \end{tabular*}
\end{ruledtabular}
\end{table*}

For $X \ll \min(N_1, N_2)$, we obtain the approximate consensus time $T_\text{sparse}$ for all three update rules (SLP, RLP, and CLP) by assuming that  $T_\text{sparse}$ can be approximated by the truncated power series in Eq.~\eqref{eq:T_sparse}.
For the SLP, we insert Eq.\ \eqref{eq:T_sparse} into Eq.\ \eqref{eq:T_SLP_for_small_X} and compare the constant terms on the left-hand and right-hand sides of the equation.
The result is
\begin{equation}
  N_1 c_{02} + N_2 c_{20} = -2 N_1 N_2.
  \label{eq:cij_VM_interior}
\end{equation}
Because the blue consensus is an absorbing state, we must have $T(0, 0) = 0$.
By inserting this condition into Eq.~\eqref{eq:T_sparse}, we obtain the condition
\begin{equation}
  16 c_{00} + 4(c_{02} + c_{11} + c_{20}) + c_{22} = 0.
  \label{eq:cij_VM_blue_cons}
\end{equation}
Apart from Eq.~\eqref{eq:cij_VM_interior} and Eq.~\eqref{eq:cij_VM_blue_cons}, there are three more conditions on the coefficients $c_{ij}$ that follow from evaluating Eq.~\eqref{eq:T_sparse} in the polarized corner $(\rho_1, \rho_2) = (1, 0)$ and along the edges of the parameter space: $(\rho_1, 0)$ with $\rho_1 \notin \{0, 1\}$ and $(0, \rho_2)$ with $\rho_2 \notin \{0, 1\}$.
We can combine all the conditions on $c_{ij}$ into the matrix equation
\[
  \mathbf{M}_\text{SLP}\,
  \mathbf{c}_\text{SLP}
  = \mathbf{v}_\text{SLP}
\]
with $\mathbf{c}_\text{SLP} = (c_{00}, c_{02}, c_{11}, c_{20}, c_{22})^\top$,
\begin{widetext}
\begin{equation*}
 \mathbf{M}_\text{SLP} =
  \begin{pmatrix}
    0 & N_1 & 0 & N_2 & 0 \\
    16 & 4 & 4 & 4 & 1 \\
    0 & -4 P_2 X & 2 (P_1 + P_2) X & -4
    P_1 X & -(P_1 + P_2) X\\
    0 & -4 P_2 X & 2 P_1 X & 4 N_2 P_1
    P_2 & N_2 P_1 P_2\\
    0 & 4 N_1 P_1 P_2 & 2 P_2 X & -4
    P_1 X & N_1 P_1 P_2
  \end{pmatrix}
  \text{, and }
  \mathbf{v}_\text{SLP} =
    -2 N_1 N_2
    \begin{pmatrix}
    1\\
    0\\
    2 P_1 P_2\\
    4 P_1 P_2\\
    4 P_1 P_2
  \end{pmatrix}.\\
\end{equation*}

The corresponding matrix equations for the other two processes, $\mathbf{M}_\text{RLP}\, \mathbf{c}_\text{RLP} = \mathbf{v}_\text{RLP}$ and $\mathbf{M}_\text{CLP}\, \mathbf{c}_\text{CLP} = \mathbf{v}_\text{CLP}$, can be found in a similar manner.
In these cases, the matrices and vectors are
\begin{align*}
 & \mathbf{M}_\text{RLP} =
    \begin{pmatrix}
    0 & N_1 & 0 & N_2 & 0 \\
    16 & 4 & 4 & 4 & 1 \\
    0 & -4 N_1^2 P_1 X & 2 (N_1^2 P_1 + N_2^2 P_2) X & -4
    N_2^2 P_2 X & -(N_1^2 P_1 + N_2^2 P_2) X\\
    0 & -4 N_1^2 P_1 X & 2 N_2^2 P_2 X & 4 N_1 N_2^2 P_1
    P_2 & N_1 N_2^2 P_1 P_2\\
    0 & 4 N_1^2 N_2 P_1 P_2 & 2 N_1^2 P_1 X & -4 N_2^2
    P_2 X & N_1^2 N_2 P_1 P_2
  \end{pmatrix},
  \;
  \mathbf{v}_\text{RLP} =
  -2 N_1 N_2
  \begin{pmatrix}
    1\\
    0\\
    2 N_1 N_2 P_1 P_2\\
    4 N_1 N_2 P_1 P_2\\
    4 N_1 N_2 P_1 P_2
  \end{pmatrix},\\
  & \mathbf{M}_\text{LD} =
  \begin{pmatrix}
    0 & N P_2 & 0 & N P_1 & 0 \\
    16 & 4 & 4 & 4 & 1 \\
    0 & -4 N N_1 X & 2 N^2 X & -4
    N N_2 X & -N^2 X\\
    0 & -4 N N_1 X & 2 N N_2 X & 4 N N_1 N_2 P_1 & N N_1 N_2 P_1\\
    0 & 4 N N_1 N_2 P_2 & 2 N N_1 X & -4
    N N_2 X & N N_1 N_2 P_2
  \end{pmatrix}
  \text{, and }
  \mathbf{v}_\text{LD} =
  -2 (N_1^2 P_1 + N_2^2 P_2)
  \begin{pmatrix}
  1\\
  0\\
  2 N_1 N_2\\
  4 N_1 N_2\\
  4 N_1 N_2
  \end{pmatrix}.
\end{align*}
\end{widetext}

It is possible to express the solutions of $\mathbf{M}_\text{SLP}\, \mathbf{c}_\text{SLP} = \mathbf{v}_\text{SLP}$, as well as the corresponding equations for the other two processes, in a closed form.
We use these exact solutions for the interpolation shown as solid curves in Fig.~\ref{fig:T}.
The expressions are long and not immediately insightful; thus, we omit them here.
However, one can easily infer the leading-order terms in the limit $\frac{X}{\min(N_1, N_2)} \to 0$ of $c_{ij}$ from the general form of the matrices $\mathbf{M}_\text{SLP}$, $\mathbf{M}_\text{RLP}$, and $\mathbf{M}_\text{CLP}$, and the vectors $\mathbf{v}_\text{SLP}$, $\mathbf{v}_\text{RLP}$, and $\mathbf{v}_\text{CLP}$.
We find
\begin{align}
& c_{11, \text{SLP}} \to -\frac{2 N_1 N_2 P_1 P_2}{(P_1 + P_2) X}\ ,
\label{eq:c11SLP}\\
& c_{11, \text{RLP}} \to -\frac{2 N_1^2 N_2^2 P_1 P_2}{(N_1^2 P_1 + N_2^2 P_2) X}\ ,\\
& c_{11, \text{CLP}} \to -\frac{2 N_1 N_2 (N_1^2 P_1 + N_2^2 P_2)}{N^2 X}\ .
\end{align}
In the special case of a polarized initial condition (i.e., $\rho_1 = 1$ and $\rho_2 = 0$), we can simplify Eq.~\eqref{eq:T_sparse} using Eq.~\eqref{eq:cij_VM_blue_cons},
\begin{align}
T_\text{sparse}(1, 0)
& = c_{00} + \frac{1}{4} \left(c_{02} - c_{11} + c_{20}\right) + \frac{1}{16}\,c_{22}
\nonumber\\
& = -\frac{1}{2}\,c_{11}.
\label{eq:T_sparse_10_from_c11}
\end{align}
The combination of Eqs.~\eqref{eq:c11SLP}--\eqref{eq:T_sparse_10_from_c11} explains Eq.~\eqref{eq:T_sparse_10} in the main text.
For the non-polarized initial state $\rho_1 = \rho_2 = 1/2$, which we consider in Appendix~\ref{app:non-polarized}, we obtain
\begin{equation}
    T_\text{sparse}\left(\frac{1}{2}, \frac{1}{2}\right) = c_{00} = -\frac{1}{4}\,c_{11}.
\label{eq:T_sparse_unpolarized_from_c11}
\end{equation}

\section{\label{app:arbitrary_X}Approximate mean consensus time for arbitrary inter-community connectivity}

If $X \gg \min(N_1, N_2)$, we observe in agent-based simulations that $\rho_1 \approx \rho_2$ after a short transient; thus, the two-dimensional parameter space $(\rho_1, \rho_2)$ effectively becomes one-dimensional.
The random variable $R(\rho_1, \rho_2)$ from Eq.~\eqref{eq:R} is a convenient choice to turn the two-dimensional input into a one-dimensional random variable because $\rho_1 \approx \rho_2$ implies $\rho_1 \approx \rho_2 \approx R(\rho_1, \rho_2)$.
Similar adiabatic approximations have also been applied in earlier studies~\cite{sood_voter_2005, sood_voter_2008, constable_population_2014, constable_fast-mode_2014, gastner_consensus_2018, gastner_voter_2019}.

With this approximation, the functions $g_1$ and $g_2$ in Table~\ref{tab:app_diff_approx} are zero, and the partial differential equation~\eqref{eq:app_diff_approx} becomes the ordinary differential equation
\begin{equation}
\frac{R (1 - R)}{J} \frac{d^2 T}{dR^2} = -1,
\label{eq:T_VM_ODE}
\end{equation}
where $J$ is a process-dependent parameter,
\begin{equation}
  J =
  \begin{cases}
    \frac{(2 E_1 + X) (2 E_2 + X)(r_{12} + r_{21})^2}
    {s_{12} + s_{21}} & \text{SLP,}\\[6pt]
     \frac{4 N_1 N_2 (E_1 + E_2 + X)^2}{N_1 (2 E_2 + X)^2 + N_2 (2 E_1 + X)^2} & \text{RLP,}\\[6pt]
     N & \text{CLP}
  \end{cases}
\end{equation}
with 
\begin{align*}
& r_{ij} = N_i^2 (2E_j + X),\\
& s_{ij} = N_i^2 (2 E_j + X)^2 [N_i^3 P_i (2 E_j + X) + N_j X (2 E_i + X)].
\end{align*}

We denote the solution to Eq.~\eqref{eq:T_VM_ODE} by $T_\text{dense}$.
The subindex ``dense'' expresses that we obtained the equation under the assumption that $X \gg \min(N_1, N_2)$.
The unique solution, subject to the absorbing boundary condition $T_\text{dense}(0) = T_\text{dense}(1) = 0$,
is
\begin{equation}
  T_\text{dense}(R) = -J [R \ln R + (1-R) \ln (1-R)].
  \label{eq:T_dense}
\end{equation}

The comparison of $T_\text{dense}$ with numerical results obtained using Monte Carlo simulations, shows that the fit is excellent if communities are densely interwoven. However, $T_\text{dense}$ is a poor fit if the inter-clique connectivity is sparse because, under this condition, there can be a long transient after the initial state during which the assumption $\rho_1 \approx \rho_2$ is invalid.
For sparse inter-clique connectivity, the approximation $T_\text{sparse}$ of Eq.~\eqref{eq:T_sparse_10_from_c11} is more accurate.
By interpolating between $T_\text{sparse}$ and $T_\text{dense}$, we can construct a function $T_\text{interp}$ that takes advantage of a better approximation in the respective parameter range.

As a first step toward the interpolation, we calculate the difference $\Delta$ between $T_\text{sparse}$ for a fixed value of $X$ and the asymptotic value of $T_\text{sparse}$ in the limit of dense inter-community connectivity,
\begin{align}
& \Delta(\rho_1, \rho_2, X)  =
\label{eq:Delta}\\
& \qquad T_\text{sparse}(\rho_1, \rho_2, X) - \lim_{X' \to \infty} T_\text{sparse}(\rho_1, \rho_2, X').
\nonumber
\end{align}
Here, we explicitly include the dependence on $X$ in the list of function arguments.
We then obtain the interpolation by adding $\Delta$ as a correction term to $T_\text{dense}$:
\begin{align}
  & T_\text{interp}(\rho_1, \rho_2, X) =\\
  & \qquad T_\text{dense}[R(\rho_1, \rho_2, X), X] + \Delta(\rho_1, \rho_2, X).
  \nonumber
\end{align}
The solid curves in Fig.~\ref{fig:T} represent $T_\text{interp}$ for different processes and parameters.
The interpolation technique was proposed by Gastner and Ishida~\cite{gastner_voter_2019} for the special case of the RLP, where both communities were assumed as cliques.
Figure~\ref{fig:T} reveals that $T_\text{interp}$ also provides a good fit for more general two-community networks and for different processes.
This analytic method allows the exploration of the parameter space more efficiently than time-consuming Monte Carlo simulations.

\end{document}